\documentclass[preprint]{aastex}

\def\degree{\ifmmode {^\circ}\else {$^\circ$}\fi}
\def\rstar{\ifmmode {\, R_{\star}}\else $R_{\star}$\fi}
\def\msol{\ifmmode {\, M_{\odot}}\else $M_{\odot}$\fi}
\def\rsol{\ifmmode {\, R_{\odot}}\else $R_{\odot}$\fi}
\def\lsol{\ifmmode {\, L_{\odot}}\else $L_{\odot}$\fi}
\def\msolyr{\ifmmode {\, M_{\odot}\,{\rm yr}^{-1}}\else $M_{\odot}\,{\rm yr}^{-1}$\fi}
\def\mdot{\ifmmode {\,\dot{M}}\else $\dot{M}$\fi}
\def\mdotyr{\ifmmode {\,\dot{M}\,yr^{-1}}\else $\dot{M}\,yr^{-1}$\fi}

\begin{document}

\title{Discovery of Five New R Coronae Borealis Stars in the MACHO Galactic Bulge Database}

\author {A. Zaniewski$^{1,2,3}$, Geoffrey C. Clayton$^{2,4}$, D. L. Welch$^{5}$, Karl D. Gordon$^{6}$, D. Minniti$^{7}$, and K. H. Cook$^{8,9}$}

\altaffiltext{1}{Dept. of Physics, West Virginia University, Morgantown, WV 26505; azaniews@mix.wvu.edu}
\altaffiltext{2}{Maria Mitchell Observatory, 4 Vestal St., Nantucket, MA 02554}
\altaffiltext{3}{Now at Astronomy Department, 601 Campbell Hall, University of California at Berkeley
Berkeley, CA 94720-341}
\altaffiltext{4}{Dept.~of Physics \& Astronomy, Louisiana State
University, Baton Rouge, LA 70803;
gclayton@fenway.phys.lsu.edu}
\altaffiltext{5}{Dept. of Physics, McMaster University, Hamilton, Ontario,  L8S
4M1 Canada; welch@physics.mcmaster.ca}
\altaffiltext{6}{Steward Observatory, University of Arizona, Tucson,
AZ 85721; kgordon@as.arizona.edu}
\altaffiltext{7}{Departimento de Astronomia, P. Universidad Catolica,
Casilla 104, Santiago 22, Chile;
dante@astro.puc.cl}
\altaffiltext{8}{Lawrence Livermore National Laboratory, Livermore, CA
94550; kcook@igpp.llnl.gov}
\altaffiltext{9}{Center for Particle Astrophysics, University of
California, Berkeley, CA 94720}

\begin{abstract}
We have identified five new R Coronae Borealis (RCB) stars in the Galactic bulge using the  MACHO Project photometry database, raising the total number of known Galactic RCB stars to about 40. We have obtained spectra to confirm the identifications.  The fact that four out of the five newly identified RCB stars are ``cool''  (T$_{eff} <$ 6000 K) rather than ``warm'' (T$_{eff} >$ 6000 K) suggests that the preponderance of warm RCB stars among the existing sample is a selection bias.  These cool RCB stars  are redder and fainter than their warm counterparts and may have been missed in surveys done with blue plates.  Based on the number of new RCB stars discovered in the MACHO bulge fields, there may be $\sim$250 RCB stars in the reddened ``exclusion" zone toward the bulge.
\end{abstract}



\section{Introduction}
R Coronae Borealis (RCB) stars are believed to be in a very short--lived evolutionary phase which is not yet well understood (Clayton 1996). Our ability to comprehend these stars depends critically on enlarging the small sample of known objects and also on understanding the severe selection biases at work.  With irregular dramatic declines of several magnitudes over only tens of days, these carbon--rich, hydrogen--deficient supergiants have proven rare, with only about 35 RCB stars known in the Galaxy (Clayton 1996).  Their spectacular declines in brightness at optical wavelengths are attributed to the ejection of material which condenses to form thick dust clouds that obscure the photosphere.
 
         Two competing theories for the evolution of RCB stars are the Double Degenerate (DD) and the Final Helium Shell Flash (FF) models.  The former involves the merger of two white dwarfs in a binary system (Iben, Tutukov, \& Yungleson\  1996a; Saio \& Jeffery 2000).  In the FF scenario, a pre--white dwarf expands to supergiant size through a helium shell flash (Iben et al. 1983; Renzini 1990).  This model suggests a relationship between RCB stars and PN; the evidence for such a relationship became stronger recently when outbursts were observed for three hot PN central stars (V4334 Sgr, FG Sge, V605 Aql) that transformed them into cool giants with the spectral properties of an RCB star (Kerber et al. 1999; Asplund et al. 1999; Clayton \& De Marco 1997; Gonzalez et al. 1998).
         
        Recently, Alcock et al. (2001) reported the discovery of eight new RCB stars in the Large Magellanic Cloud (LMC) using MACHO Project photometry. Five RCB stars have been  discovered in the SMC (Tisserand et al. 2004; Morgan et al. 2003). Recently identified Galactic RCB stars include Z UMi (Benson et al. 1994), ES Aql (Clayton et al. 2002) and V2552 Oph (Hesselbach et al. 2003; Rao \& Lambert 2003). Most of the newly discovered stars are cool (T$_{eff} <$ 6000 K) RCB stars, implying that observational bias and not population may have precluded their discovery until now. Lawson \& Cottrell (1990) discussed an exclusion zone towards Galactic center (GC), where excessive reddening prevented the discovery of Galactic bulge RCB stars in previous studies, such as the Harvard Observatory plate survey which had a limiting magnitude of B$\sim$12. 
Alcock et al. (2001) suggest that M$_V \propto$ T$_{eff}$. So, cool, reddened RCB stars may have been systematically missed in previous surveys. 
This paper reports the results of a  search of the MACHO Project database for new RCB stars in the Galactic bulge.

\section{Observations}

The MACHO Project (Alcock et al. 1992) was designed to search for gravitational microlensing events.
 It collected two--color photometric data on millions of stars in the Large and Small Magellanic Clouds, and the Galactic bulge.  
The survey spanned seven  years, from 1992 to 1999, and utilized a dedicated 50-inch telescope at Mt. Stromlo, Australia.  The camera was mounted at prime focus, had a field of view of 0.5 deg$^2$, and was custom built for this project (Stubbs et al. 1993).  Two--color photometry in blue ( $B_{macho}$, $\sim$4400--5900 \AA) and red ( $R_{macho}$, $\sim$5900--7800 \AA) light was obtained simultaneously
by using a dichroic beam splitter.
The camera system incorporated eight 2048x2048 pixel Loral CCDs, with 2x2 mosaics in each focal plane.  The 15 \micron~pixels mapped to 0\farcs63 on the sky. The photometric data were reduced using SODOPHOT,
 a profile--fitting photometry routine derived from DoPHOT (Schecter, Mateo, \& Saha 1993). Galactic
 bulge data were obtained for 94 fields, covering approximately 50 deg$^{2}$.
The database was searched for RCB star candidates as described in Section 3. MACHO can detect stars down to about 20th magnitude and is saturated for stars brighter than about 12th magnitude. 
        
We utilized the calibration relation of Popowski et al. (2003) (based on Alcock et 
al. 1999) to convert to Johnson V and Kron--Cousins R  bandpasses:\\
V = 23.70 + 0.82 $B_{MACHO}$  + 0.18 $R_{MACHO}$\\
R = 23.41 + 0.18 $B_{MACHO}$  + 0.82 $R_{MACHO}$.\\
The MACHO light curves, calibrated in this way for the new RCB candidate stars are shown in Figure 1; $(V-R)_{KC}$ colors are also plotted.  The stars are listed in Table 1 using the standard ``field.tile.sequence" number, which designates a particular object in the database, and their J2000.0 equatorial coordinates.  Finder charts are shown in Figure 2.

We obtained JHK photometry from the Two Micron All Sky Survey (2MASS) project for each of the stars. The data are listed in Table 2 with the Modified Julian Dates of the observations. The typical 1-$\sigma$ 2MASS errors are $\pm$ 0.03 mag in all three bands.

Spectroscopic observations for all of the stars except 118.18666.100 were obtained in 2003 May from Las Campanas observatory, Chile, with the Low--Dispersion Survey Spectrograph (LDSS) mounted on the 6.5m Magellan II telescope.  The grating used gives a resolution of $\sim$10 \AA, and a useful range of about 3600-7000 \AA.  The slits used were long enough to simultaneously record background sky and object, allowing accurate sky subtraction for each star.  The spectrograph was wavelength calibrated using two arc lamps, and flux calibration was obtained by observing one standard star. Images were also acquired at Magellan in order to check the star IDs.
Spectra for 118.18666.100 were obtained on 2003 July 7 using the Steward Observatory's 2.3m Bok telescope at Kitt Peak, Tucson, with the B\&C spectrograph. The grating gives a resolution of $\sim$5 \AA, and a useful range of about 3200-8000 \AA.
The spectra of the candidate stars are shown in Figure 3.
Twelve additional candidate stars were rejected on the basis of spectra obtained at Steward and Las Campanas. Three of these stars belong to the AM Her class of magnetic cataclysmic variables (Hellier 2001). The lightcurve of an AM Her star, featuring active and inactive states, looks very similar to an RCB lightcurve. Few AM Her stars are identified in the optical (Drissen et al. 1994) so searching for RCB-like lightcurve behavior seems to be an efficient method of discovering these rare variables. The remainder of the rejected candidates are K and M stars. They are listed in Table 3.

\section{New RCB Stars in the Galactic Bulge}
The signature of an RCB star is its lightcurve (Clayton 1996). A star may stay at maximum brightness for months or years, then  show unusual sharp declines at irregular intervals. The declines happen quickly, with a drop of 3 magnitudes or more taking a few days or weeks. Recovery typically lasts months or years (Payne--Gaposchkin \& Gaposchkin 1938). The RCB stars are cool (T$_{eff}$ = 5000-7000 K), hydrogen-deficient carbon stars, as revealed by their spectra. The hydrogen lines are weak at best, as is the G-band (CH)\footnote{V854 Cen, one of the most active RCB stars, shows fairly
strong Balmer lines and CH band
(Kilkenny \& Marang
1989; Lawson \& Cottrell 1989).}.
RCB stars exhibit strong carbon lines and molecular bands of CN and C$_2$ that dominate their spectra. However, they typically have little or no $^{13}$C.
RCB stars  often exhibit regular or semi--regular pulsations with $\Delta$V of a few tenths of a magnitude and periods of 40-100 days.  They typically also have an infrared excess.

Three previously known RCB stars lie within the MACHO Galactic bulge fields. These are V739 Sgr (123.24377.5), V3795 Sgr (161.2445.2938) and VZ Sgr (117.25166.4791); however, these stars do not lie within the top fields, meaning there was very little data taken for them.  GU Sgr lies just outside the area covered by the MACHO fields.

We searched the MACHO database for stars with large declines in brightness, and further culled candidates by excluding stars whose declines were periodic.  
The seven year MACHO lightcurves for these candidates were then examined visually to identify likely RCB stars.
We used spectroscopic data to confirm the final candidates as RCB stars or to otherwise classify them.  All of our newly-discovered RCB stars are present in sky surveys, such as 2MASS, but have not previously been identified as variable stars.
We applied the criteria used by Alcock et al. (2001) to identify the candidates as RCB stars.
The rate of decline, dm/dt (magnitudes/day), quantifies the characteristic sharp decline, and the lightcurves were examined by eye for evidence of pulsations.
The spectra were examined for evidence of hydrogen by looking at the Balmer lines and the G band of CH at 4300 \AA. The presence of $^{13}$C was searched for in the isotopic bands of C$_2$ and CN. In particular, the Swan bands, $^{13}$C$^{12}$C and $^{13}$C$^{13}$C near 4700 \AA, other C$_2$ bands in the 6000-6200 \AA\ region, and the $^{13}$CN band near 6250 \AA\ were examined.
The photometric and spectroscopic characteristics are listed in Table 4.


One star, 118.18666.100, shows similar spectral features to other warm RCB stars, such as W Men,
although the S/N of the spectra were not sufficient for a definite identification. The typical low-resolution, warm RCB spectrum is almost featureless since there are no Balmer lines and the molecular features are very weak.
 The lightcurve of 118.18666.100 shows one sharp decline of $>$ 2 mag followed by a slow return to maximum.  It experiences pulsations similar to other RCB stars.

The four other stars, 135.27132.51, 301.45783.9, 308.38099.66, and 401.48170.2237 show spectra typical of cool RCB stars with very strong molecular absorption bands of C$_2$ and CN.
Two of the stars, 401.48170.2237 and 301.45783.9 are very active, showing several deep declines. The lightcurve coverage isn't as complete for 308.38099.66 and 135.27132.51 but the former shows
evidence for two declines and the latter shows evidence for one. Pulsations can be seen in the lightcurves of 401.48170.2237, 301.45783.9, and 308.38099.66.  The fragmentary lightcurve of 135.27132.51 does not permit a search for pulsations.

\section{IR Excesses}
All of our newly-discovered RCB stars have been observed in the infrared by the 2MASS survey. The 2MASS data are listed in Table 2.
Their positions in the H-K/J-H color-color plot would fall right on those observed for S Aps, a cool RCB star. See Figure 7 of Feast (1997).
The 2MASS data only give a snapshot  of each star. One star, 118.18666.100, was at or near maximum light when observed; 301.38099.66 was recovering from a decline; 308.38099.66 was beginning a sharp decline; and 401.48170.2237 was in a deep decline. The three stars in decline lie above and to the right of the stars at maximum light following the behavior seen for other RCB stars (Feast 1997). Due to the fragmentary nature of the 135.27132.51 lightcurve, we don't know where it was when observed by 2MASS. But from its position in the H-K/J-H color-color plot, it was not in a decline.

\section{The Population of RCB stars}

Recently, Alcock et al.(2001) estimated 3200 Galactic RCB stars, based upon the number of observed RCB stars in the LMC and extrapolating to the Galaxy. Both of the evolutionary theories referenced in the introduction, the Double Degenerate and the Final Helium Shell Flash models, suggest that the RCB stars are an old population (Clayton 1996). The Double Degenerate scenario would suggest a population of about 1000 Galactic RCB stars (Webbink 1984).  Iben, Tutokov \& Yungleston (1996b) put the Galactic RCB population resulting from the same scenario to be about 300 stars, and they calculate that the Final Flash scenario would imply anywhere from 30 to 2000 RCB stars at any given time.  All of the evidence thus far suggests that there are many more than the 40 known RCB stars in the Galaxy.

It is a significant extrapolation from the current small sample to estimate the population of Galactic RCB stars. There are some severe selection biases affecting the known sample. Most of the RCB stars were discovered on plate surveys that went down to B$\sim$12. 
Discovery is also hampered by the fact that the RCB stars start at maximum and then get fainter so the maximum light brightness of the star has to be significantly above the plate limit to be detected as an RCB star. 
In the LMC where a reasonably unbiased census was possible, Alcock et al. (2001) found that M$_V \propto$ T$_{eff}$ such that M$_V$ falls from -5 to -3 as T$_{eff}$ goes from $\sim$7000 to 5000 K. They also found that the cool (T$_{eff} <$ 6000 K) RCB stars far outnumber the warm (T$_{eff} >$ 6000 K) RCB stars. 
 Of the Galactic RCB stars identified before 1988, primarily in plate surveys, that have T$_{eff}$ estimates, 16 are warm and 4 are cool. Starting with the fluke discovery of V854 Cen in 1988\footnote{It is V=7.1 at maximum light but had spent most of the 20th century in a deep decline.}, thirteen additional RCB stars have been identified. Of these, ten are cool and three are warm, just the opposite of the sample discovered earlier.  These new RCB stars have a cool/warm ratio similar to that seen in the LMC. Therefore, cool RCB stars may have been heavily undercounted in previous Galactic surveys because they are redder and fainter than their warmer counterparts.
The selection effects are exacerbated in the direction toward the GC where reddening is high. 

With the addition of the five new RCB stars reported in this work, there are now about 40 RCB stars known in the Galaxy. Their distribution on the sky is shown in Figure 4.  This larger sample, for which we now have more accurate estimates of T$_{eff}$ and M$_V$, allow us to significantly improve on the distributions attempted previously (Lawson \& Cottrell 1990; Drilling 1986). The new stars do not show a significantly different distribution from those discovered earlier. It is very hard to discern the nature of the population of the RCB stars in the Galaxy based on such a small sample.  There is no indication of a difference in distribution for the warm and cool RCB stars.  In Figure 4b, we have plotted the ``exclusion zone" postulated by Lawson \& Cottrell (1990) where RCB stars may have been missed due to large amounts of dust extinction. The addition of newly discovered stars plus additional stars having T$_{eff}$ estimates have filled in the nearby portion of the zone but the actual distribution of the RCB stars remains mysterious.

The distribution on the sky and radial velocities of the RCB stars tend toward those of the bulge 
population (Drilling 1986; Jeffery et al. 1987). 
In the discussion following Drilling's paper, 
it is reported that new data show that the radial velocities of the RCB stars are significantly 
lower than those of the EHe stars. The EHe stars are hotter than the RCB stars and don't make dust. The scale heights for these two groups of stars have 
been derived as z = 1700 pc for the EHe stars and 400 pc for the RCB stars (Iben 
\& Tutukov 1985). So while the EHe stars seem to be bulge/Population II stars, the RCB 
stars may be more like old disk/Population I stars. 
Figure 4 shows that while the RCB stars are concentrated toward the GC in the sky, most of them lie 2-5 kpc from the Sun and 3-5 kpc from the GC. One can easily believe that the exclusion zone could be populated with very faint reddened RCB stars. But if there is a spheroidal distribution of RCB stars 5 kpc in radius around the GC, then there should be large numbers of lightly reddened RCB stars, with V mags brighter than 12, lying above and below the plane of the Galaxy. See Figure 4b. This is not seen. 
Taking into account the selection bias constraints outlined above, the distribution of RCB stars is consistent with a thick disk population which has a scale height of  $\sim$750 pc and a scalelength of 2.8 kpc (Robin et al. 1996). As shown in Figure 5, the distribution of the RCB stars is also similar
 to the distant PNe (Drilling 1986). See Figure II-1 of Pottasch (1984) for comparison. There is a sharp peak in the density of RCB stars around the GC very much as there is in the PNe consistent with a Bulge population. However, the distant PNe are thought to lie close ($<$ 1 kpc) to the GC and as shown in Figure 4, the known RCB stars are actually quite far from the GC, nearer to the Sun.

Of the 94 MACHO Bulge fields, only a fraction had consistently good coverage over the seven years. For each MACHO field, we estimated the fraction of the mission with coverage (as opposed to the percentage of nights observed). Using this method, we estimated that thirty-three fields had coverage of less than 10\% and twenty-seven fields had coverage equal to 40\%.  Of the five stars that we discovered, four lie in fields with  40\% coverage. 
Any RCB stars brighter than V$\sim$12 will not be detected in the MACHO data because they will be saturated on the CCD. This is probably a very small number.
Alcock et al. (2001) estimated the detection rate for finding RCB stars in the MACHO database to be 
75\%, based upon the AAVSO monitoring of 31 known RCBs. Of these, 23 experienced a decline of greater than three magnitudes during the seven year time span of the MACHO project. 
Given that each MACHO Bulge field contains 0.5 deg$^2$, we are surveying $\sim$13.5 deg$^2$ in the top 27 fields in MACHO Bulge database. This is less than 3\% of the area of the Lawson \& Cotrell (1990) exclusion zone. Based on the 5 new stars identified and assuming a 75\% success rate, there would be $\sim$250 RCB stars in the exclusion zone to a depth of about 7 kpc, the distance to the farthest RCB star discovered in this study.

\section{Summary}

Our search of the MACHO bulge photometry database for new RCB stars had the following results.

\noindent
 $\bullet$ Five new RCB stars were discovered bringing
the total known in the Galaxy to $\sim$40.\\
$\bullet$ The discovery ratio of 4:1 for cool:warm RCB stars in this study agrees with the predominance of cool RCB stars found in the LMC, and indicates a selection bias for previously discovered RCB stars.\\
$\bullet$ The true spatial distribution of RCB stars is still uncertain, although the projected maps indicate a possible thick disk population.\\
$\bullet$ We estimate the population of RCB stars in the reddened exclusion zone toward the Galactic Center to be about 250.\\

To determine the true distribution and population of RCB stars in the Galaxy, MACHO-like surveys will have to be done of other patches of sky at different Galactic latitudes and longitudes.

\acknowledgments

This project was supported by the NSF/REU grant AST-0097694 and the Nantucket Maria Mitchell Association. We would also like to thank Vladimir Strelnitski, Director, Maria Mitchell Observatory. D.M. is supported by FONDAP Center for Astrophysics No. 15010003. D. L. W. was supported by the Natural Sciences and Engineering Research Council of Canada (NSERC).
This publication makes use of data products from the Two Micron All Sky Survey, which is a joint project of the University of Massachusetts and the Infrared Processing and Analysis Center/California Institute of Technology, funded by the NASA and NSF.
KHC's work was performed under the auspices of the U.S. Department of
Energy, National Nuclear Security Administration by the University of
California, Lawrence Livermore National Laboratory under contract No.
W-7405-Eng-48.


\clearpage

\clearpage

\begin{figure*}
\figurenum{1}
\includegraphics*[width=6in]{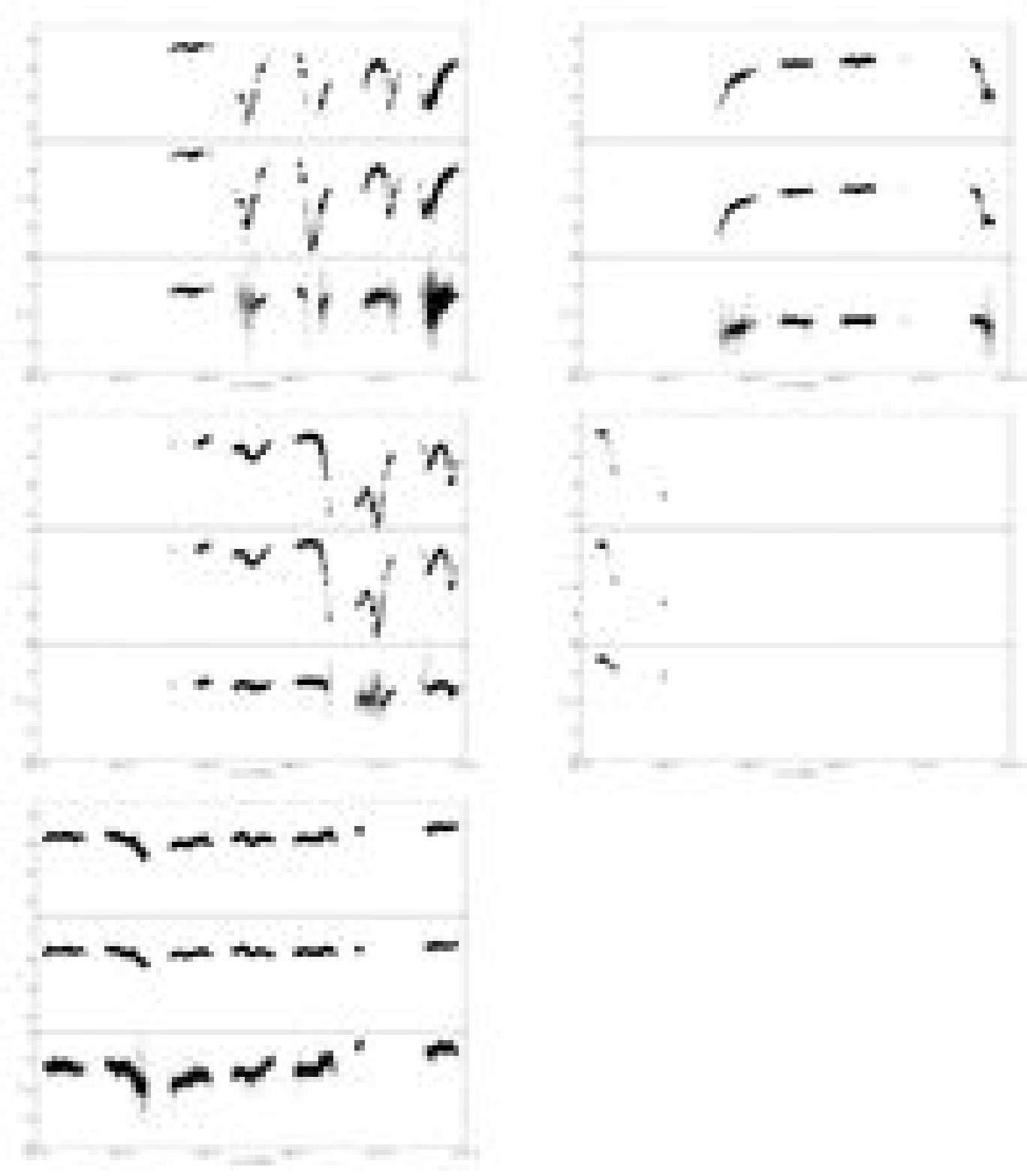}
\caption{MACHO Project lightcurves for confirmed RCB stars 301.45783.9 (upper left),308.38099.66 (upper right), 401.48170.2237 (middle left),135.27132.51 (middle right), and 118.18666.100 (bottom). }
\end{figure*}


\begin{figure*}
\figurenum{2}
\includegraphics*[width=6in]{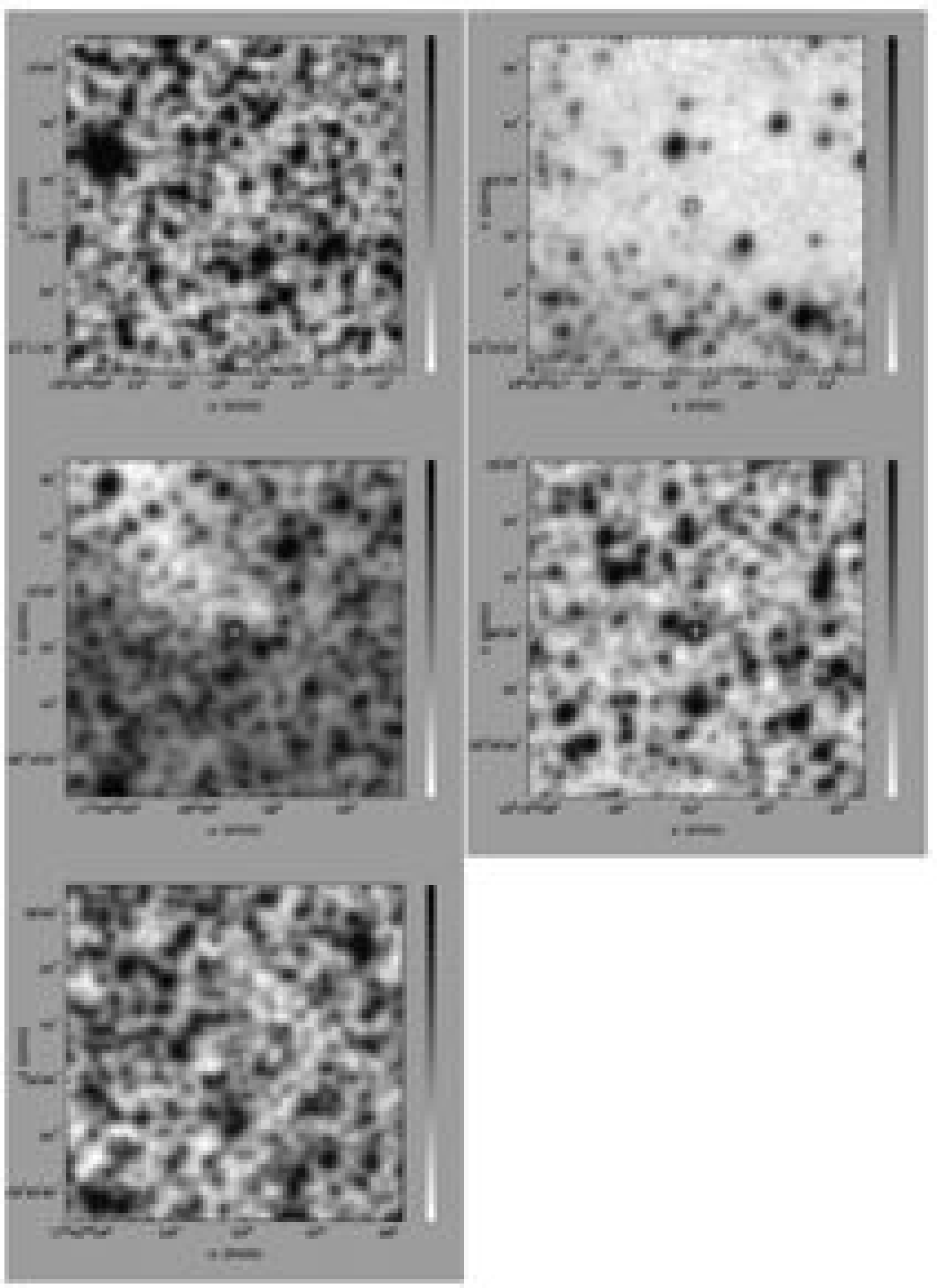}
\caption{Finding charts for RCB stars, 301.45783.9 (upper left),308.38099.66 (upper right), 401.48170.2237 (middle left),135.27132.51 (middle right), and 118.18666.100 (bottom). }

\end{figure*}


\begin{figure*}
\figurenum{3}
\includegraphics*[width=3.25in]{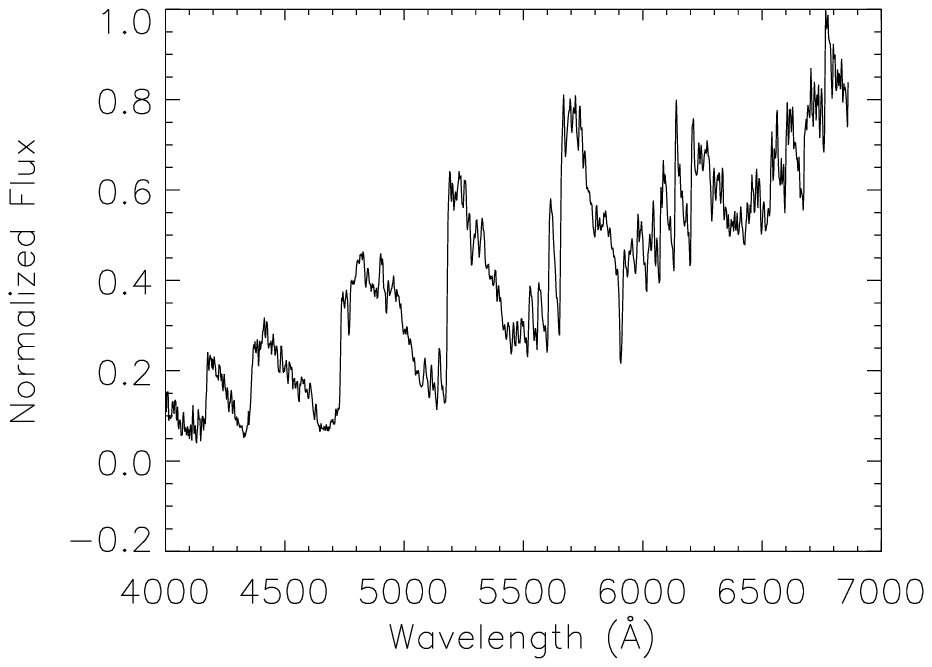}
\includegraphics*[width=3.25in]{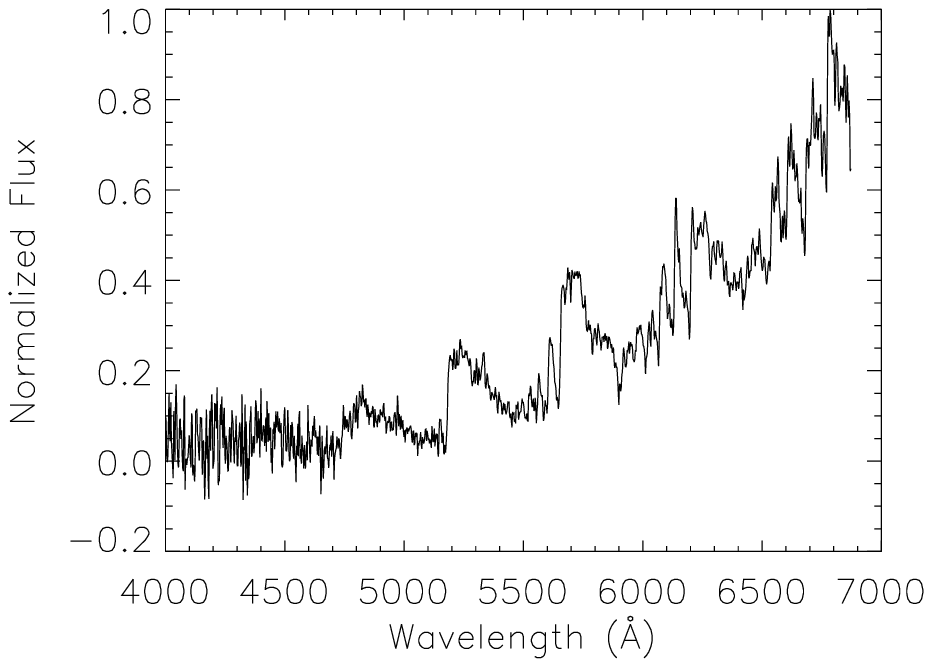}
\includegraphics*[width=3.25in]{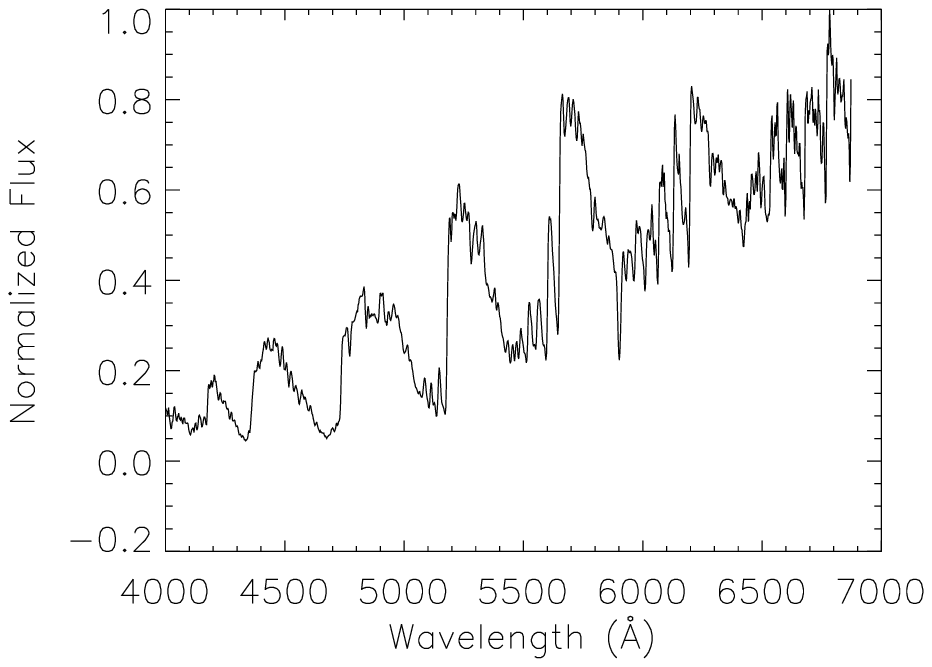}
\includegraphics*[width=3.25in]{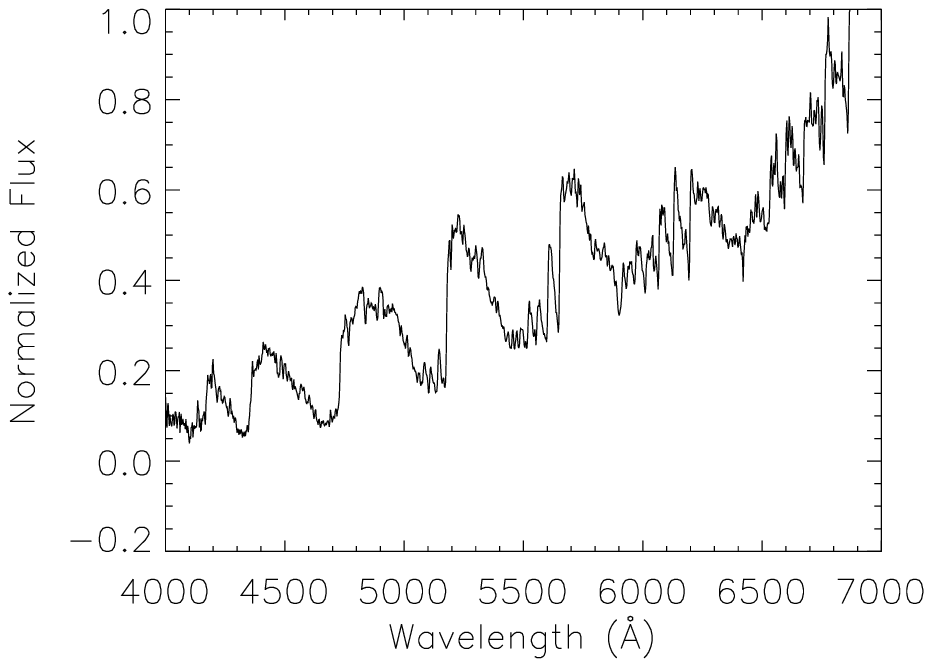}
\includegraphics*[width=3.25in]{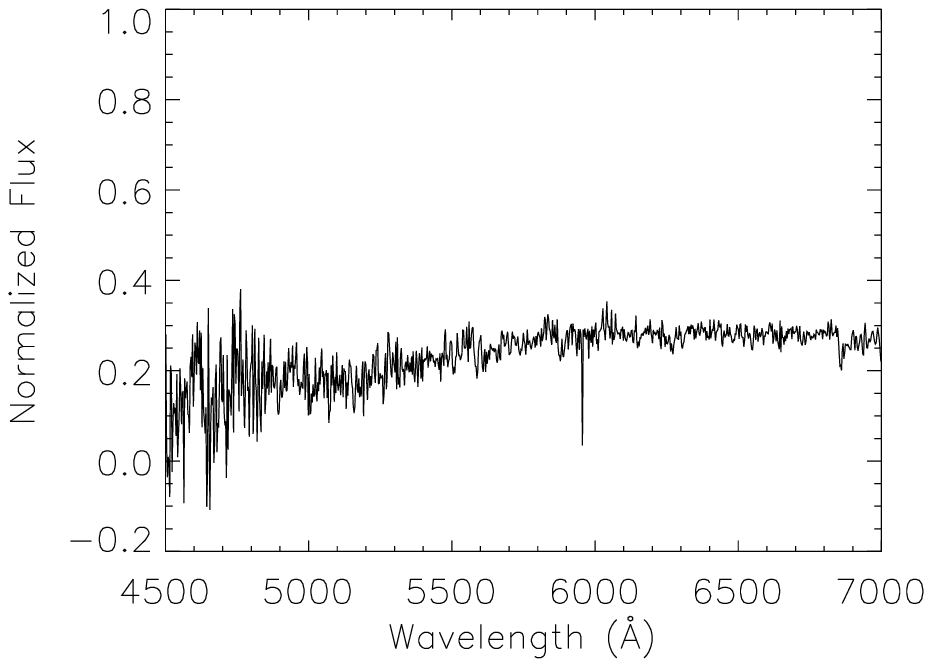}
\includegraphics*[width=3.25in]{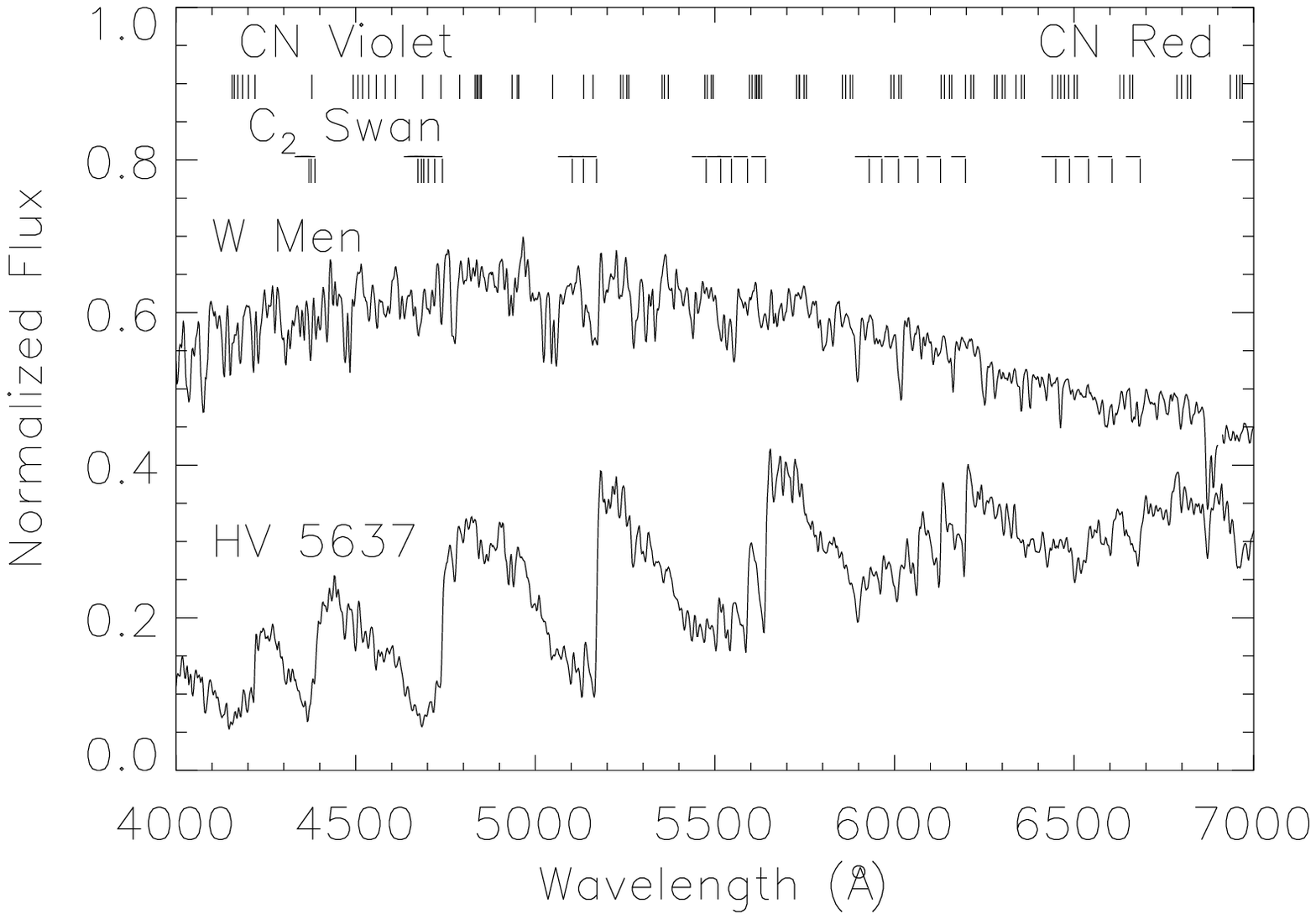}
\caption{Spectra of confirmed cool RCB stars, 301.45783.9 (upper left),308.38099.66 (upper right), 401.48170.2237 (middle left), and 135.27132.51 (middle right). These spectra show strong absorption bands of C$_2$ and CN (Alcock et al. 2001). Spectrum of the confirmed warm RCB star 118.18666.100 (bottom left). This spectrum shows no strong features. Spectra of typical warm (W Men) and cool (HV 5637) RCB stars with molecular bands marked for comparison (bottom right) (Alcock et al. 2001).}
\end{figure*}

\begin{figure*}
\figurenum{4}
\includegraphics*[width=3.25in]{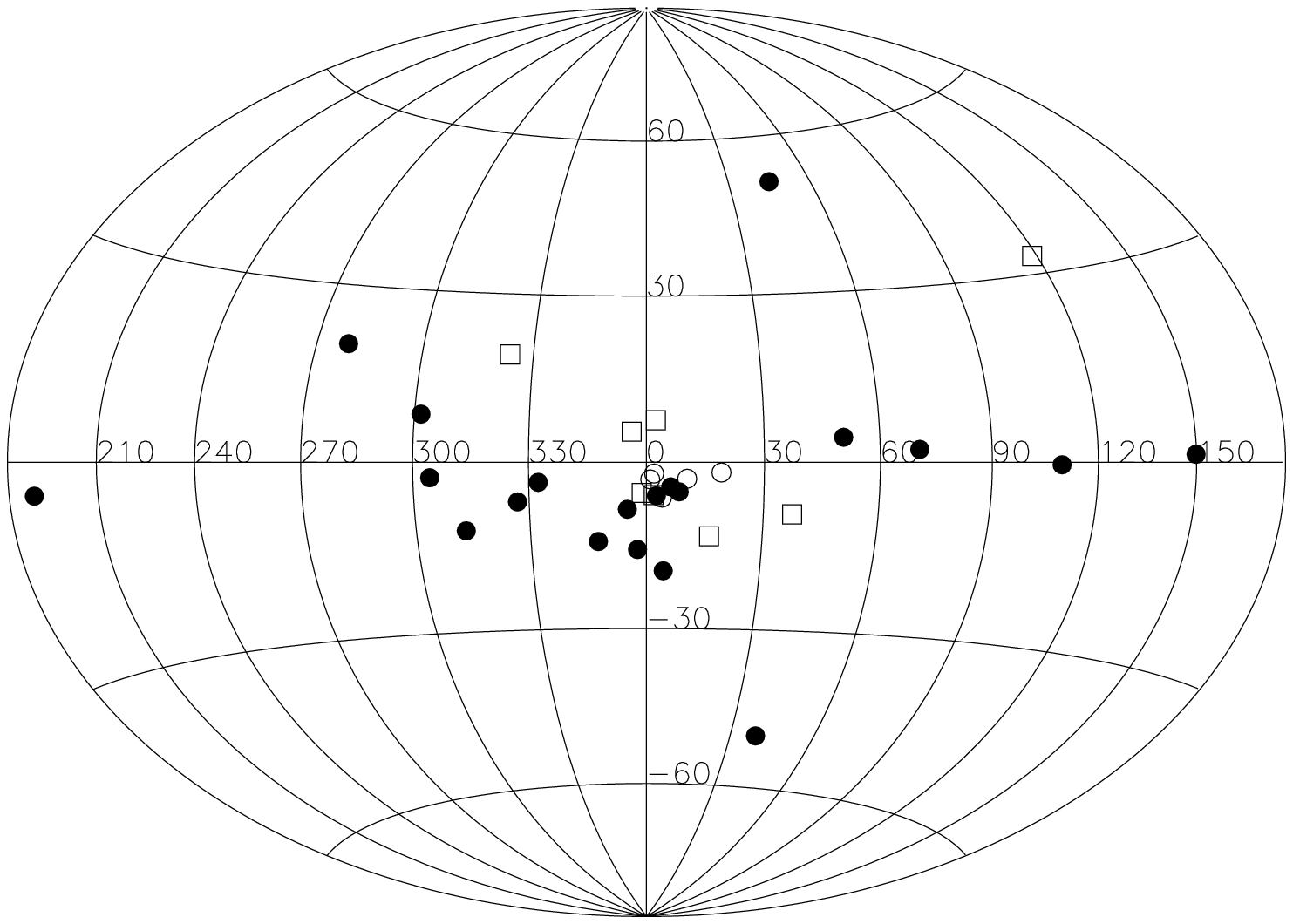}
\includegraphics*[width=3.25in,height=3.00in]{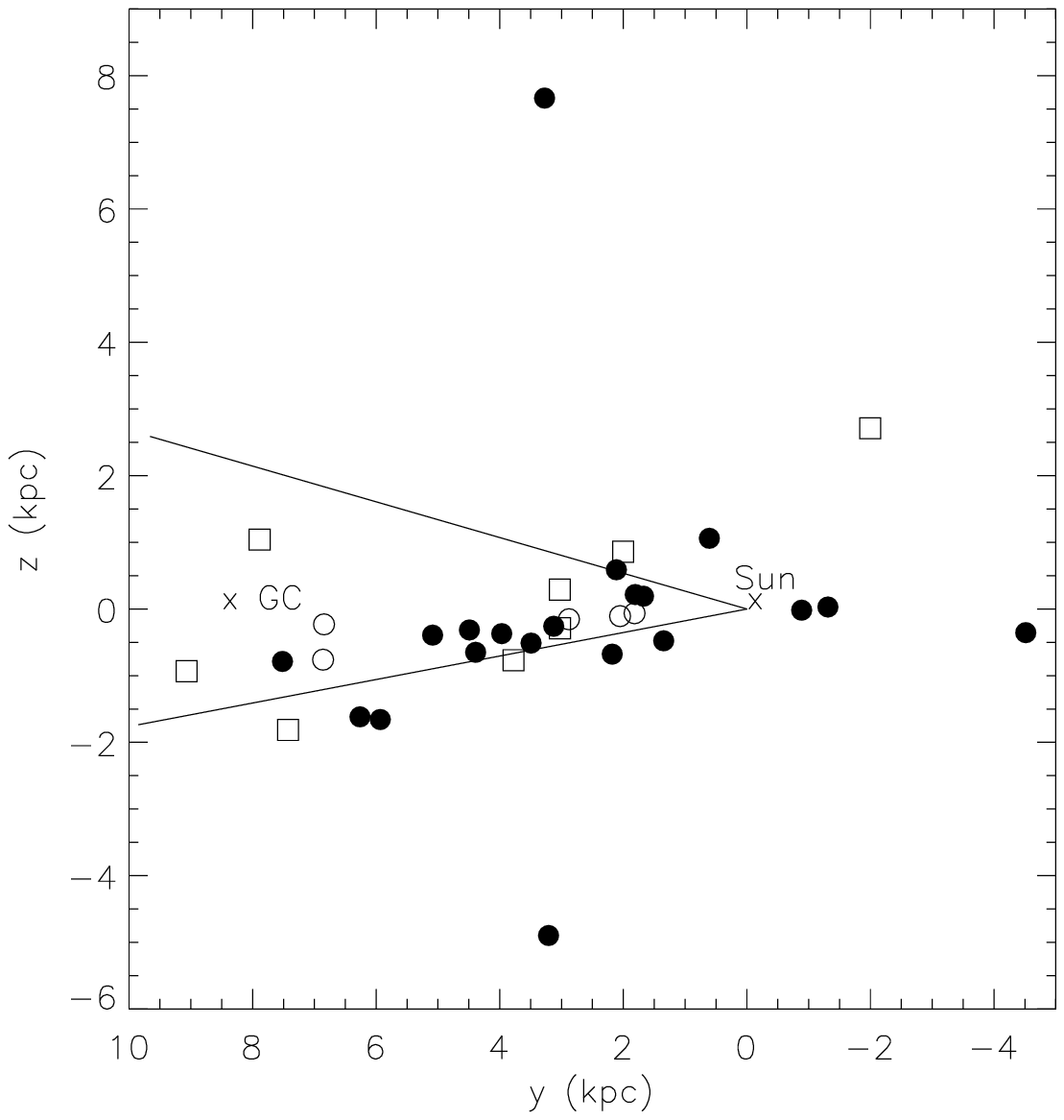}
\includegraphics*[width=3.25in,height=3.00in]{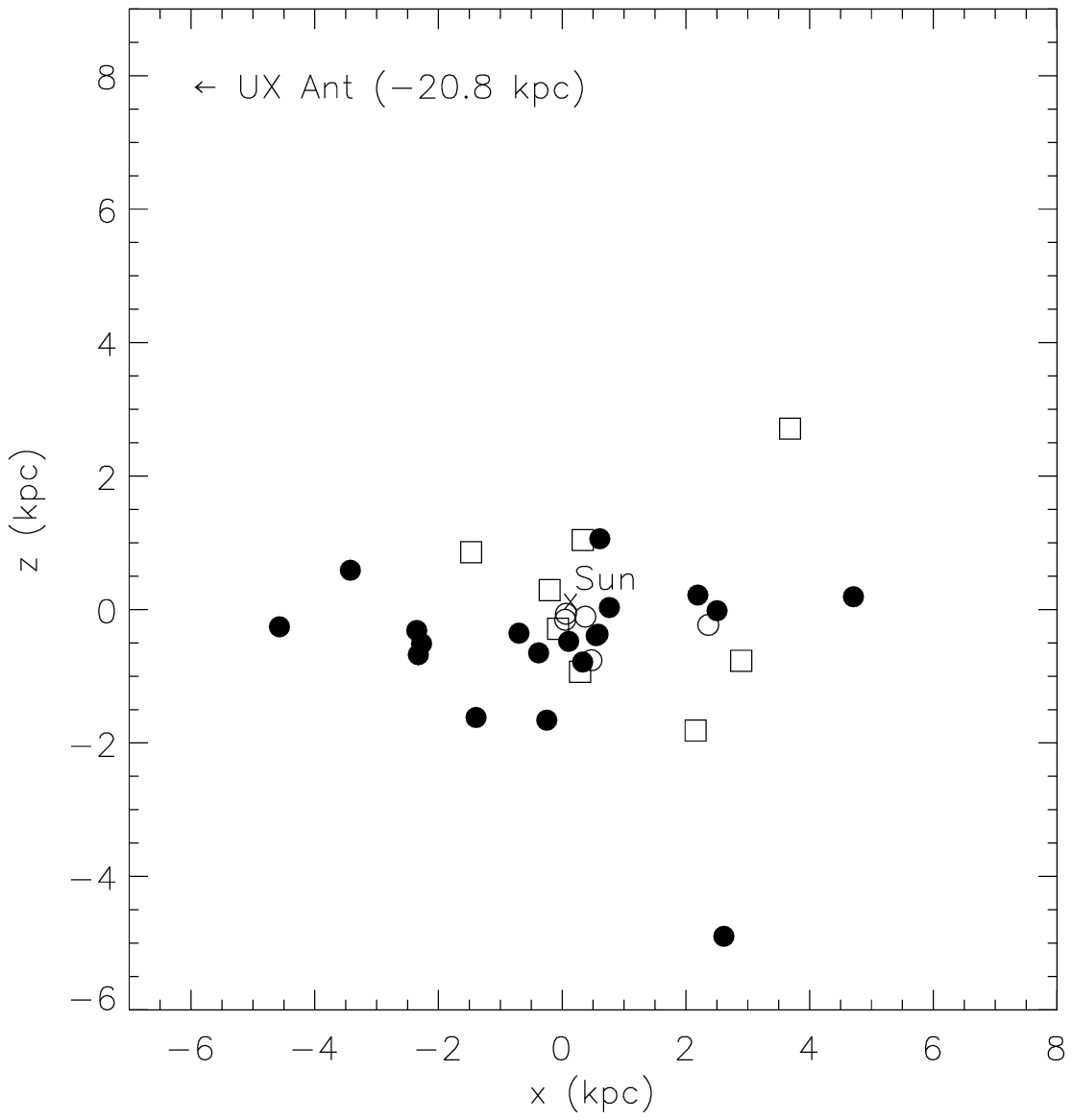}
\includegraphics*[width=3.25in,height=3.00in]{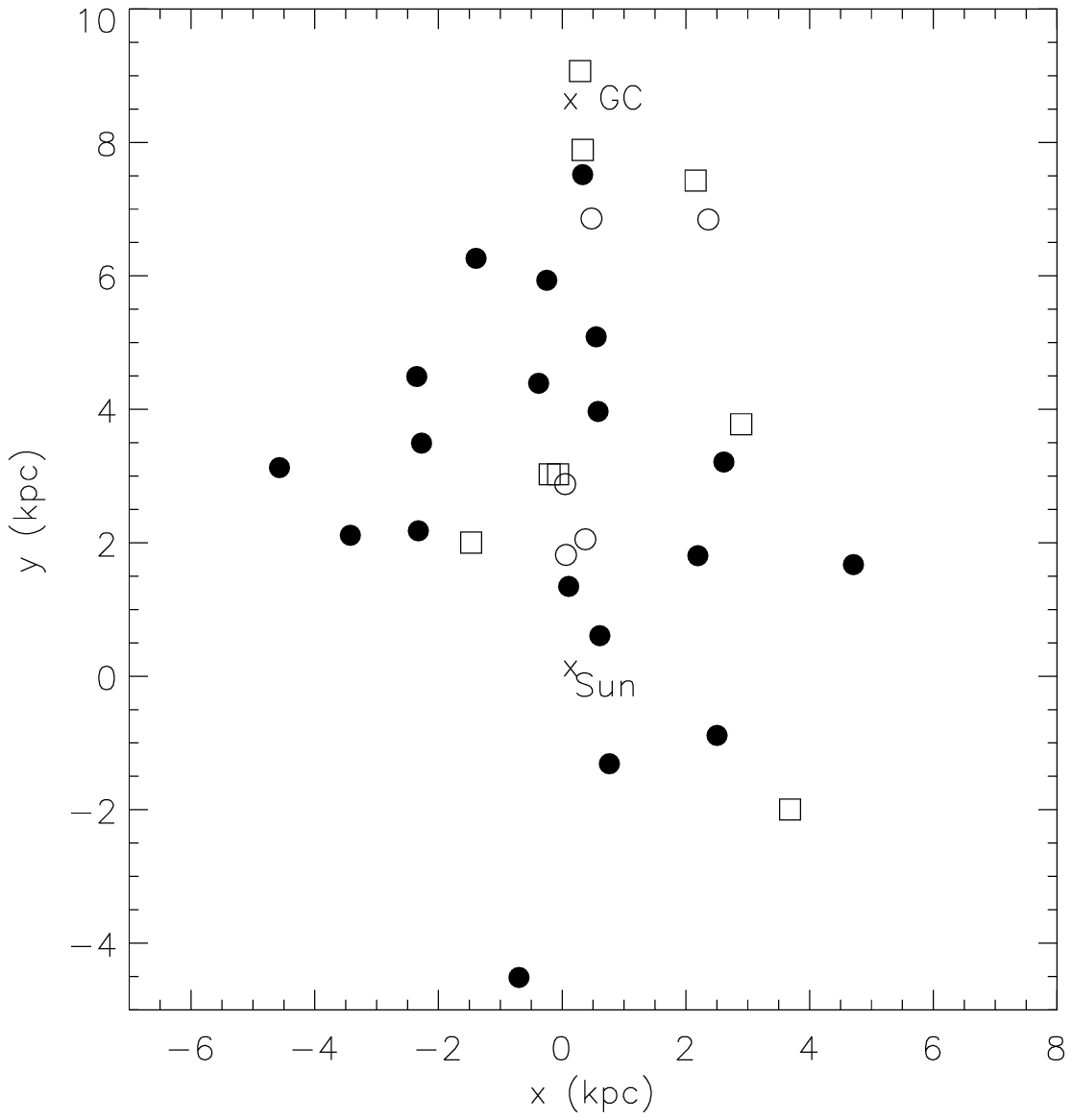}

\caption{Upper Left panel: Aitoff projection showing the distribution of RCB stars on the sky. Open circles are new MACHO Bulge RCB stars, filled circles are RCB stars identified before 1988 and open squares are RCB stars identified since 1988. Upper right, lower left and lower right panels show y vs z, x vs z and x vs y distributions of the RCB stars in units of kpc. The symbols are the same as in the upper left panel. 
The locations of the Sun and the Galactic center (GC) are marked. In the upper right panel, the solid lines show the exclusion zone from Lawson \& Cottrell (1990) at 
b=+15$\degree$ and -10$\degree$ in the direction of the GC.}
\end{figure*}

\begin{figure*}
\figurenum{5}
\includegraphics*[width=3.25in]{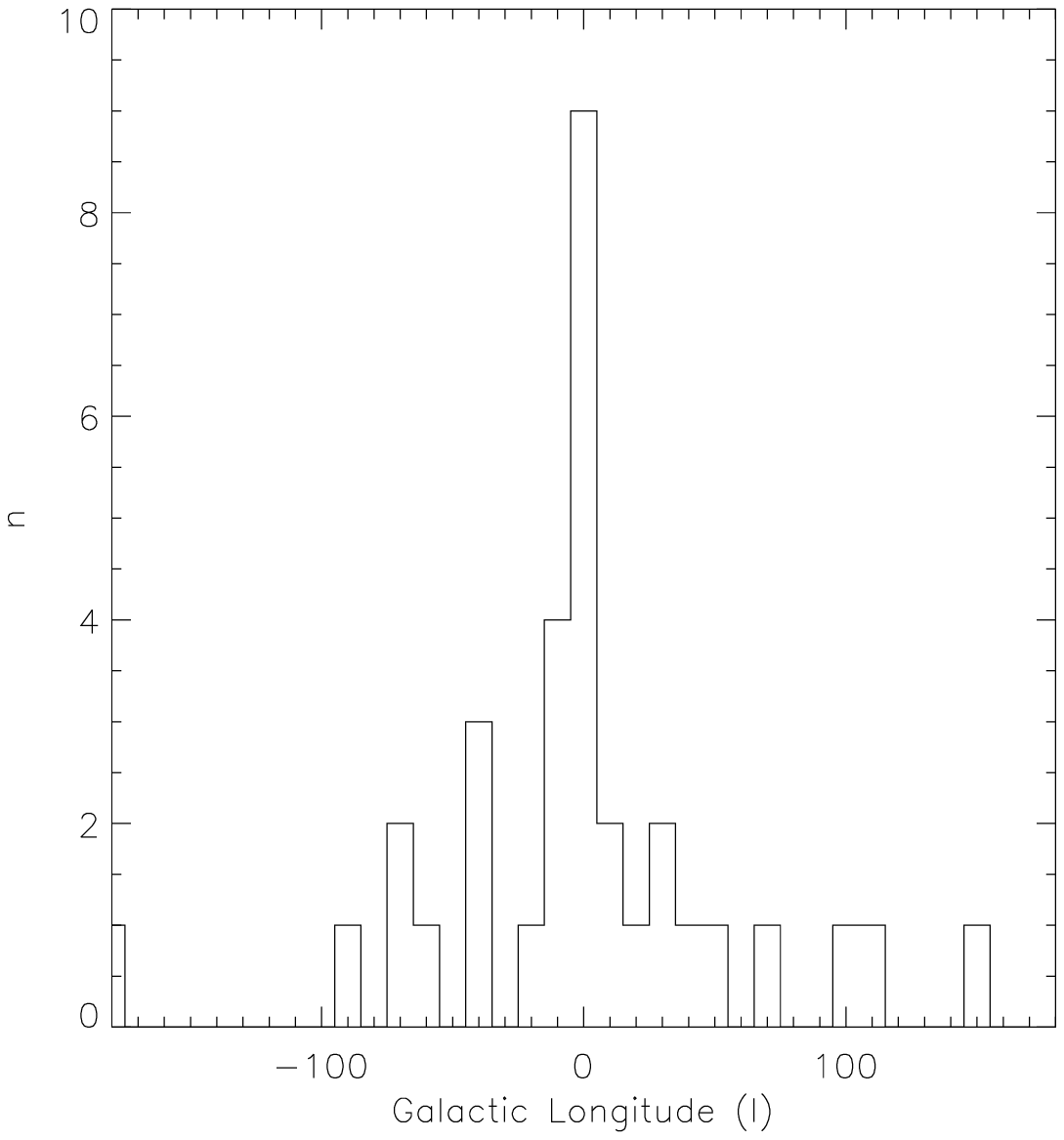}
\includegraphics*[width=3.25in]{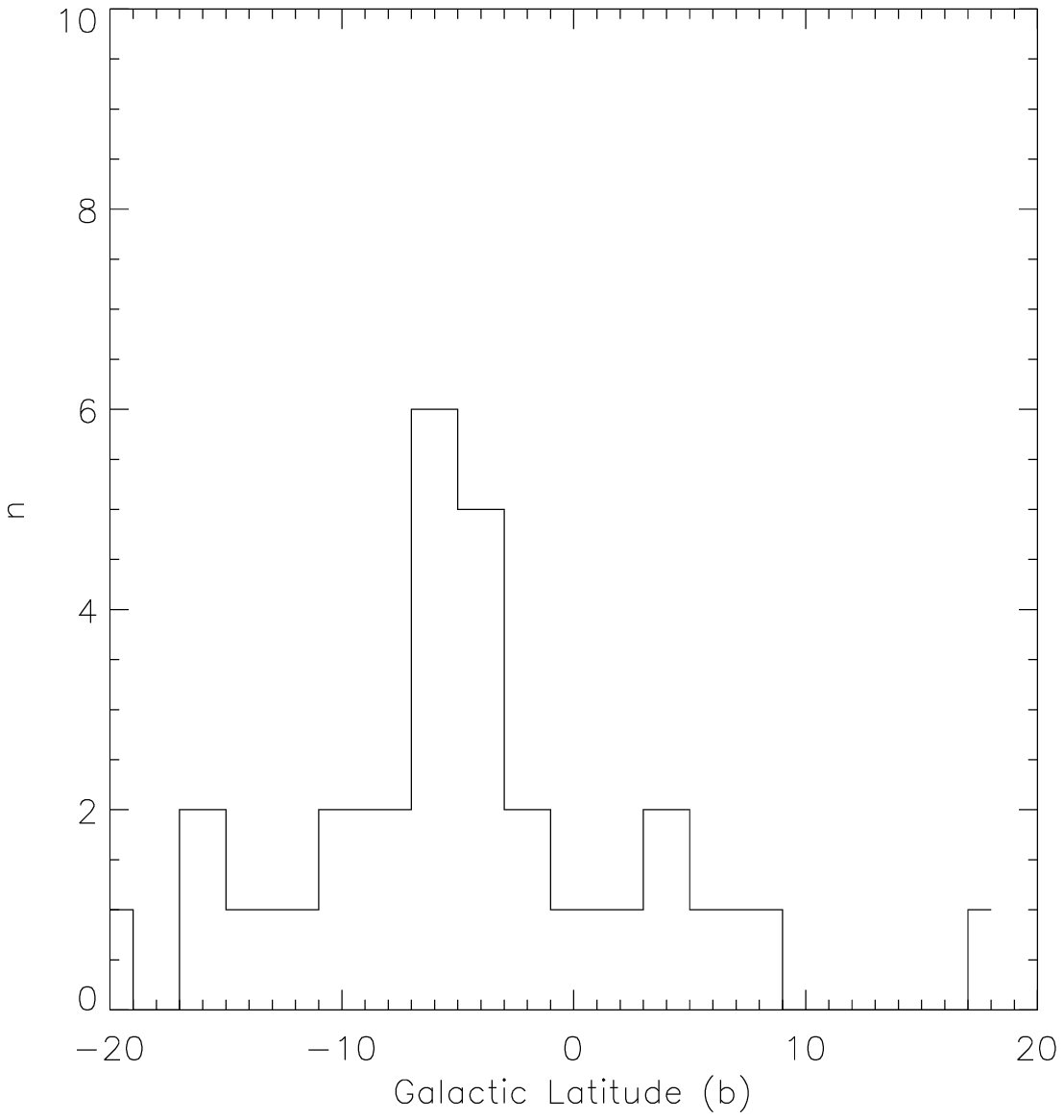}
\caption{Histogram distribution of RCB stars with Galactic latitude and longitude. See text.}
\end{figure*}

\clearpage

\begin{deluxetable}{llll}
\tablecaption{MACHO Galactic Bulge RCB Stars }
\tablenum{1}
\tablehead{\colhead{Name\tablenotemark{a}} & \colhead{MACHO Name}& \colhead{$\alpha(2000)$}&\colhead{$\delta$(2000)}}
\startdata
\underline{Known RCB Stars}&&\\
V739 Sgr (123.24377.5)&MACHO*18:13:13.1-30:15:50&18 13 13.11 &-30 15 50.4\\
V3795 Sgr (161.24445.2938)&MACHO*18:13:24.2-25:46:43&18 13 24.23  &-25 46 43.3\\
VZ Sgr (117.25166.4791)&MACHO*18:15:09.1-29:42:29&18 15 09.13  &-29 42 28.8\\
&&&\\
\underline{Confirmed New RCB Stars}&&\\
118.18666.100&MACHO*17:59:52.2:-29:39:50&17 59 52.23  &-29 39 50.0\\
135.27132.51 &MACHO*18:19:33.9-28:35:58&18 19 33.87  &-28 35 57.8\\
301.45783.9 &MACHO*18:32:18.6-13:10:49&18 32 18.60 &-13 10 48.9\\
308.38099.66&MACHO*18:19:27.4-21:24:08&18 19 27.36 &  -21 24 08.2\\
401.48170.2237&MACHO*17:57:59.0-28:18:13&17 57 59.02&-28 18 13.1  \\
\enddata
\tablenotetext{a}{Includes MACHO F.T.S. number.}
\end{deluxetable}

\begin{deluxetable}{lllll}
\tablecaption{2MASS JHK Photometry}
\tablenum{2}
\tablehead{\colhead{Name}&\colhead{MJD-2400000}&\colhead{J (mag)}&\colhead{H (mag)}&\colhead{K (mag)}}
\startdata
118.18666.100&51010&\phn9.2&\phn8.1&7.5\\
135.27132.51&51407&10.4&\phn9.7&8.9\\
301.45783.9&51366&12.2&11.0&9.9\\
308.38099.66&51339&11.9&10.1&8.6\\
401.48170.2237&51010&11.5&\phn9.8&8.4\\
\enddata
\end{deluxetable}


\begin{deluxetable}{lllllllllll}
\tablecaption{Final Identification}
\tablenum{3}
\tablehead{\colhead{$Name$}& \colhead{$V_{max}$}& \colhead{$R_{max}$}&\colhead{(V-R)}&\colhead{E(B-V)}&
\colhead{$\Delta$m}&\colhead{$\Delta$t}&\colhead{dm/dt}&\colhead{$^{13}$C}&\colhead{H/CH}&\colhead{Pulsations}}
\startdata
118.18666.100  &16.6  &14.9 &1.7 &3.0  &$>$1.5   &25   &0.06  &None &None  &yes\\
135.27132.51    &14.3  &13.1  &1.2  &1.0 &$>$3.5    &75     &0.047  &None &None         &?\\
308.38099.66    &17.3   &15.2       &2.1 &2.8    &$>$5     &\nodata     &\nodata     &None   &None &yes\\
301.45783.9     &16.3 &14.8 &1.5  &1.6&5      &110     &0.05          &None   &None           &yes\\
401.48170.2237&14.5    &12.8  &1.7  &2.0   &6      &100    &0.06   &None   &None &yes\\
\enddata
\end{deluxetable}


\begin{deluxetable}{lllll}
\tablecaption{Non-RCB Stars}
\tablenum{4}
\tablehead{\colhead{$Name$}& \colhead{$V_{max}$}& \colhead{$R_{max}$}&\colhead{(V-R)}& \colhead{Probable Classification}}
\startdata
159.25614.43    &15.2  &13.9  &1.4   &M\\
122.22949.21    &15.1   &13.7   &1.4    &M\\            
154.29038.66    &16.4  &16.2  &0.2     &AM Her\\
149.27752.4439  &16.9  &16.7    &0.2    &AM Her\\
311.37557.169   &15.3 &15.1 &0.2   &AM Her\\
304.36408.33    &17.4 &15.0 &2.4   &K Star?\\
305.36243.133  &16.6   &16.1 &0.5    &?\\
104.20900.3659  &12.5    &13.5  &1.0    &K star\\
102.26243.24    &13.6  &12.5 &1.1 &M\\
303.44915.18    &15.8 &14.1 &1.7  &M\\
102.23507.2965  &10.2 &\nodata &\nodata &M\\
148.26072.30    &14.2 &13.1 &1.1 &M\\
104.20385.189 &17.5 &15.5 &2  &M\\
\enddata
\end{deluxetable}

\end{document}